\documentclass[prb,aps,  showpacs, reprint, preprintnumbers,amsmath,amssymb,superscriptaddress,floatfix]{revtex4-1}
\usepackage[]{graphicx}      % Include figure files
\usepackage{bm}             	% bold math
\usepackage{times}			% Times font
\usepackage{tabularx}
\usepackage{color}
\usepackage{dcolumn}		% Align table columns on decimal point
\usepackage{amsmath}
\usepackage{amssymb}
\usepackage{amsfonts}
\usepackage{times}
\usepackage{tabularx}
\usepackage{xcolor,colortbl}
\newcommand{\IEF}{Institut d'Electronique Fondamentale, CNRS, Univ. Paris-Sud, Universit\'e Paris-Saclay, 91405 Orsay, France}
\newcommand{\SAMSUNG}{SAMSUNG Electronics Corporation, 601 McCarthy Blvd Milpitas, CA 95035, USA}

\begin{document}
\title{Size-dependence of nanosecond-scale spin-torque switching in perpendicular magnetized tunnel junctions}

\author{T. Devolder}
\email{thibaut.devolder@u-psud.fr}
\author{A. Le Goff} 
\affiliation{\IEF}
\author{V. Nikitin}  
\affiliation{\SAMSUNG}

\date{\today}                                           
%%%%%%%%%%%%%%%%%%%%%%%%%%%%%%%%%%%%%%%%
%
%       Abstract
%
%%%%%%%%%%%%%%%%%%%%%%%%%%%%%%%%%%%%%%%%
\begin{abstract}
We have time-resolved the spin-transfer-torque (STT)-induced switching in perpendicularly magnetized tunnel junctions (pMTJ) of diameters from 50 to 250 nm in the sub-threshold thermally activated regime. 
When the field and the spin-torque concur to both favor the P to AP transition, the reversal yields monotonic resistance ramps that can be interpreted as a domain wall propagation through the device at velocities of the order of 17 to 30 nm/ns; smaller cells switch faster, and proportionnally to their diameter. At the largest sizes, transient domain wall pinning can occasionally occur. When the field hinders the P to AP transition triggered by the spin-torque, the P to AP switching is preceded by repetitive switching attempts, during which the resistance transiently increases until successful reversal occurs. At 50 nm, the P to AP switching proceeds reproducibly in 3 ns, with a monotonic featureless increase of the device resistance.
In the reverse transition (AP to P), the variability of thermally activated reversal is not restricted to stochastic variations of incubation delays before the onset of reversal: several reversal paths are possible even in the smallest perpendicularly magnetized junctions. Besides, the non uniform nature of the magnetic response seems still present at the nanoscale, with sometimes electrical signatures of strong disorder during the AP to P reversal. The AP to P transition is preceded by a strong instability of the AP states in devices larger than 100 nm. The resistance becomes extremely agitated before switching to P in a path yielding a slow (20 to 50 ns) and irregular increase of the conductance with substantial event-to-event variability. Unreversed bubbles of typical diameter 60 nm can persist a few additional microseconds in the largest junctions. The complexity of the AP to P switching is reduced but not suppressed when the junctions are downsized below 60 nm. The instability of the initial AP state is no longer detected but the other features are maintained. In the smallest junctions (50 nm) we occasionally observe much faster (sub-1 ns) AP to P switching events that could result from a macrospin process. We discuss the origin of the switching asymmetry and the size dependence, with an emphasis on the role of the non uniformities of the stray field emanating from the reference layers of the tunnel junction, which affects the zones in which nucleation is favored.
\end{abstract}

\keywords{Switching speed, Magnetic Tunnel Junction, Perpendicular Magnetic Anisotropy, Spin torque, magnetic random access memories.}

\maketitle

%%%%%%%%%%%%%%%%%%%%%%%%%%%%%%%%%%%%%%%%
%
%                Paper
%
%%%%%%%%%%%%%%%%%%%%%%%%%%%%%%%%%%%%%%%%

The spin-transfer-torque (STT) manipulation of the magnetization in magnetic tunnel junctions (MTJ) is a very active research field, because the interplay between magnetization-dependent transport properties \cite{theodonis_anomalous_2006} and the spin torques results in a rich variety of phenomena \cite{chappert_emergence_2007}.  Perpendicular Magnetic Anisotropy (PMA) systems are an ideal playground to explore STT-induced dynamics \cite{sun_spin-current_2000, ikeda_perpendicular-anisotropy_2010}, because the high symmetry of their magnetic properties matches with the symmetries of the physical system when the nanomagnets are shaped to circular disks. Besides, high quality PMA MTJs can now be found \cite{worledge_spin_2011, gajek_spin_2012, swerts_beol_2015}, as they were optimized owing to their applications in information technologies. 

Despite their ubiquitous presence, the mechanism of STT-induced reversal in PMA MTJs is still rather secretive, and the simplicity promised by the high system symmetry is not on the cards \cite{devolder_time-resolved_2016}. The macrospin approximation used in the early models \cite{sun_spin-current_2000} has sometimes been claimed valid \cite{tomita_unified_2013, timopheev_respective_2015} but it is most of the time said to be irrelevant in practical size regime \cite{thomas_quantifying_2015, sato_properties_2014, munira_calculation_2015, chaves-oflynn_thermal_2015}, i.e. for nanomagnet disks of diameters above 20 nm. The experimental indications of the switching process are mostly indirect, since they come either from quasi-static behaviors \cite{mangin_current-induced_2006} or at best from switching probability measurements after the application of nanosecond-scale current pulses  \cite{kanai_magnetization_2014, tomita_unified_2013, worledge_spin_2011, heindl_validity_2011}. In these latter experiments, one compares the experimental and the modeled switching probability versus pulse duration curves to sort among the possible  switching scenario. However these methods cannot determine the wealth of what is really happening: single-shot time-resolved measurements are required to understand the dynamics. Single shot measurements need to cover the relevant timescales of magnetization dynamics, i.e. from dc to microwave frequencies. Such experiments were done in the past for in-plane magnetized system \cite{devolder_single-shot_2008, herault_nanosecond_2009, lacoste_modulating_2014}, however their implementation in PMA systems are scarce \cite{devolder_time-resolved_2016} and are so far restricted to large junctions, in which the reasonably low impedances facilitate the characterizations. The dramatic changes at small sizes inferred from quasi-static experiments \cite{chenchen_size_2012} are unfortunately still to be explained. 

In contrast to experiments, micromagnetic simulations may be considered as comparatively easy. However as micromagnetic simulation are difficult in sub-threshold conditions in which switching can require long waiting times and thermal activation plays a large role, most of the reported micromagnetic simulations are done for very large (hence accelerating) spin torque amplitudes that are difficult to achieve in practice because of junction fragility. Simulations often predict sub-ns switching \cite{zhang_micromagnetic_2010}, unfortunately without support or confirmation from experimental data. Besides, the magnetic constants needed for simulations -- magnetization, anisotropy, damping and exchange-- are known with insufficient accuracy as they can substantially differ from their bulk counterpart \cite{sato_effect_2016, devolder_exchange_2016}. As a result, neither the present experimental data in hand, nor the micromagnetic simulations have been conclusive on the exact switching mechanism is STT-induced dynamics in PMA MTJ nanopillars.

In this article, we report and analyze single-shot time-resolved measurements of ns-scale STT switching events in PMA MTJs. Our method relies on the measurement of the time evolution of the device conductance in sub-threshold STT conditions. We conduct this study for nanopillar diameters spanning from 250 nm and observe how the behavior gets progressively more simple as the diameter is shrunk to below 50 nm. We account for the main features of the electrical signature of the switching using two simple scenarios that are distinct for the back and forth reversal events, in line with the observed strong switching asymmetry in both switching speed and switching mode. 

The paper is organized as follows. We first briefly describe our samples, our measurement apparatus and its limits in section \ref{methods}.  The size dependence of their quasi-static properties are then reviewed in section \ref{QSproperties}. The time-resolved behaviors are then sorted according to the junction sizes, from the complex behavior of junction in the 200 nm range (section \ref{large}), to the sizes around 100 nm (section \ref{medium}) and finally to the near-50 nm cases (section \ref{small}). A discussion of the switching paths terminates the study (section \ref{discussion}).

%%%%%%%%%%%%%%%%%%%%%%%%%%%%%%%%%%%%%%%%%%%%%%%%%%%%%%%%%%%%
\section{Sample and methods} \label{methods}
%\subsection{Samples}
Our tunnel junctions are composed of a FeCoB-based free layer with dual MgO encapsulation \cite{sato_Perpendicular-anisotropy_2012}, and a hard reference system based on a well compensated synthetic antiferromagnet following standard recipes (see Fig. 1c in ref.\onlinecite{gottwald_scalable_2015}). The tunnel magnetoresistance (TMR) is typically 110\% at 500 mV of voltage bias when the device undergoes spin-torque induced switching. The TMR reaches 250\% at 10 mV of bias in the largest devices at room temperature. The stack resistance-area product is $\textrm{RA}=12~ \Omega.\mu \textrm{m}^2$. %formerly RA=10, so diameters will have to be increased by 11% \\
The full thickness of the PMA MTJs was etched into circular pillars with nominal diameters $2a$ from the sub-50 nm range to dimensions reaching 250 nm. The measured device resistances were found to be strongly correlated with their nominal size, but the overall dependence suggests an enlargement of 7 to 12 nm of the diameter upon processing. The junctions are circular, so in the remainder of this study we will assume an exact junction radius defined by $a = \sqrt{ RA / (\pi R_p)}$ where $R_p$ is the resistance of the parallel state. This leads to typical resistances of $150~\Omega$ for large junctions ($2a=250$ nm).  The small junctions are very resistive, with up to 6 and $20 ~\mathrm{k}\Omega$ in the parallel (P) and antiparallel (AP) states for the smallest junction ($2a\approx 50$ nm) that could be investigated thoroughly.

\begin{figure*}
\includegraphics[width=18 cm]{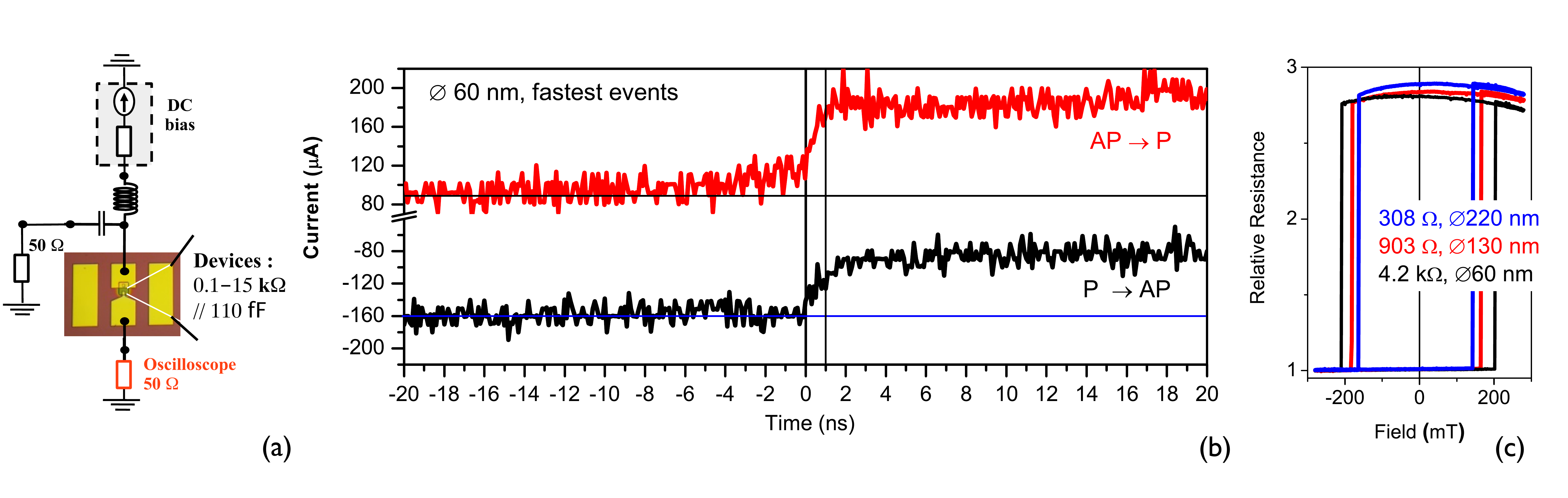}
\caption{(Color online). Panel (a): simplified sketch of the experimental set-up. The signal is the voltage delivered to a $50~\Omega$ oscilloscope by the current flowing into the device when submitted to a slow (dc) voltage ramp and coupled to charge-pump capacitor loaded by a $50~\Omega$ matching system. Panel (b): fastest observed switching events for junctions of diameter 60 nm and resistance  of $4.2~\textrm{k}\Omega$ in the parallel state. The two vertical bars are separated by one ns. The AP to P switching event is measured to last less than 1 ns, however the assessment of its exact duration is impeded by the capacitive part of the device that filters out any faster evolution. Panel (c): quasi-static resistance versus field hysteresis loops for circular devices of nominal sizes 50, 120 and 200 nm. The actual sizes are deduced from the measured device resistance in the remanent parallel state.}
\label{C5_PAPandAPPandSETUP}
\label{RHloops}
\end{figure*}

%\subsection{Electrical measurements}
The MTJ is inserted in series between ground-signal-ground coplanar waveguides (Fig.~\ref{C5_PAPandAPPandSETUP}). The device integration is designed to ensure a minimal capacitance by minimizing the area of the face-to-face surfaces of the top an bottom electrodes. Using a 44 GHz Vector Network Analyzer with on-chip full 2-port Short Open Load Through calibration, we measured the scattering matrix of the device. Its behavior could be correctly accounted for (not shown) by a capacitance of $C=110~\textrm{fF}$ in parallel with the (resistive) MTJ.  This parasitic capacitance is independent of the MTJ size; it originates from the areas of the facing top and bottom electrodes in the surroundings of the nanopillar. The time resolution of our measurements is thus bounded by the $RC$ product of the junctions. As a result of the large TMR, we thus expect a better time resolution near the P state than near the AP state. Our worse time-resolution is expected to be 1 ns for the smallest junctions, while sub-ns resolution is granted for our largest junctions. In the smallest junctions, the fastest switching events can be of sub-nanosecond duration [Fig.~\ref{C5_PAPandAPPandSETUP}(b)] such that the details of such switching events cannot be recorded accurately.

For switching experiments, the sample were characterized in a set-up whose essential features are described in Fig.~\ref{C5_PAPandAPPandSETUP}(a): a kHz-rate $2~\mathrm{V^{pp}}$ triangular voltage ramp is applied to the sample, which delivers its current to a $50~\Omega$ oscilloscope connected in series. While we ramp the applied voltage, we capture the electrical signature of magnetization switching by measuring the voltage delivered by the device to a high bandwidth $50~\Omega$ oscilloscope. The time origins in the time-resolved experiments is defined by this capturing event. This measurement procedure implies that the studied reversal regime is the sub-threshold thermally activated reversal switching. We have chosen not to perform analog amplification to avoid frequency-dependent distortions of the signals that appeared to complicate the understanding \cite{devolder_time-resolved_2016}. The drawback is that the signal-to-noise ratio of our measurement degrades as the inverse of the device area. The trigger input of the oscilloscope is fed by a high-pass filtered replica of the signal, while the trigger is gated by the voltage polarity to select either of the switching transition during the voltage ramp. The oscilloscope bandwidth was set to 6 GHz to avoid capturing noise above the $RC$ bandwidth of the sample. When specified -- notably for the junctions of diameter 50 nm for which a substantial degradation of signal-to-noise ratio is inevitable-- a mathematical smoothing was done to reduce the noise equivalent bandwidth to 0.6 GHz.

%%
%	Figure
%%
%

%%%%%%%%%%%%%%%%%%%%%%%%%%%%%%%%%%%%%%%%%%%%%%%%%%%%%%%%%%%%
\section{Quasi-static properties} \label{QSproperties}
\begin{figure}
\includegraphics[width=8 cm]{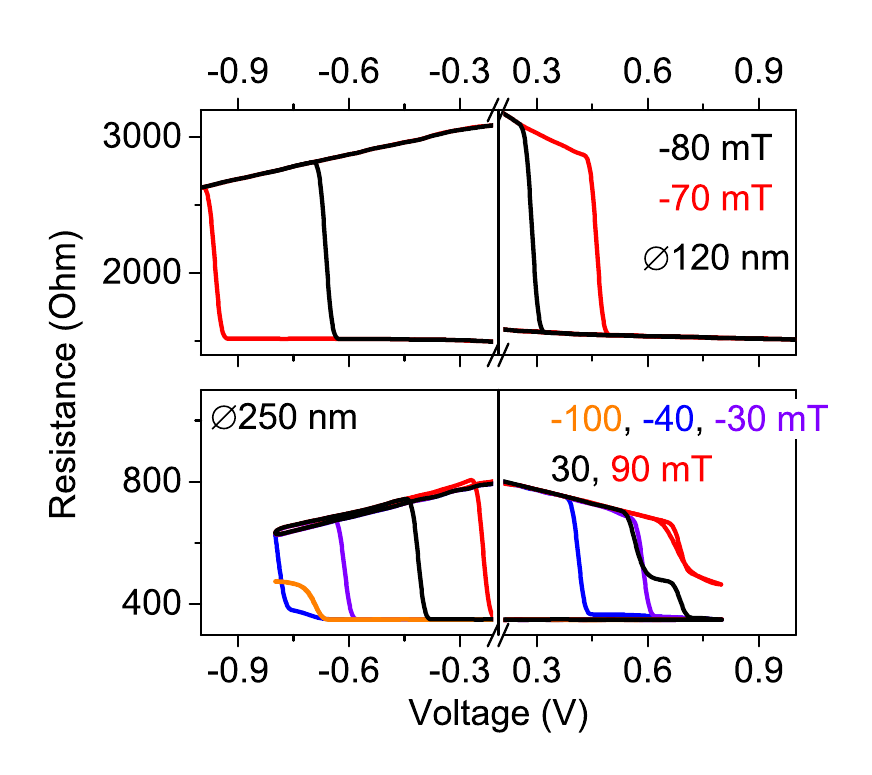}
\caption{(Color online) Quasi-static resistance versus voltage hysteresis loops for circular devices of sizes 120 nm (top panel) and 250 nm (bottom panel) for various applied fields. Notice the plateaus at intermediate resistance level at high fields for the largest junction.} % C9 and C12 devices, nouvelles tailles prises en compte .
\label{RVloops}
\end{figure}

The quasi-static properties of our samples are illustrated in Figs.~\ref{RHloops}(c) and \ref{RVloops}. Resistance versus field minor loops indicate that the device coercivity $\mu_0 H_c$ increases from 0.14 to 0.23 T when reducing the device size down to our smallest size. Part of this coercivity increase reflects the change of effective anisotropy $H_k$ due to the decrease of demagnetizing effects linked to the change of the device aspect ratio upon diameter reduction.  
  
Our convention is that positive fields favor the AP configuration [Fig.~\ref{RHloops}(c)]. The loop miscentering indicates that the dipolar coupling with the reference system changes from AP coupling (+26 mT) % c12: 26 mT ; C12: 25 mT; C9: 9 mT; C7: 15 mT
for the largest junctions to weak P coupling for the smallest junctions, but the coupling never exceeds 26 mT. The small and size-dependent field offset will be referred later as the compensation field, or as the neutral field.

%%
%	Figure
%%
%

%%
%	Figure
%%
%
\begin{figure}
\includegraphics[width=8 cm]{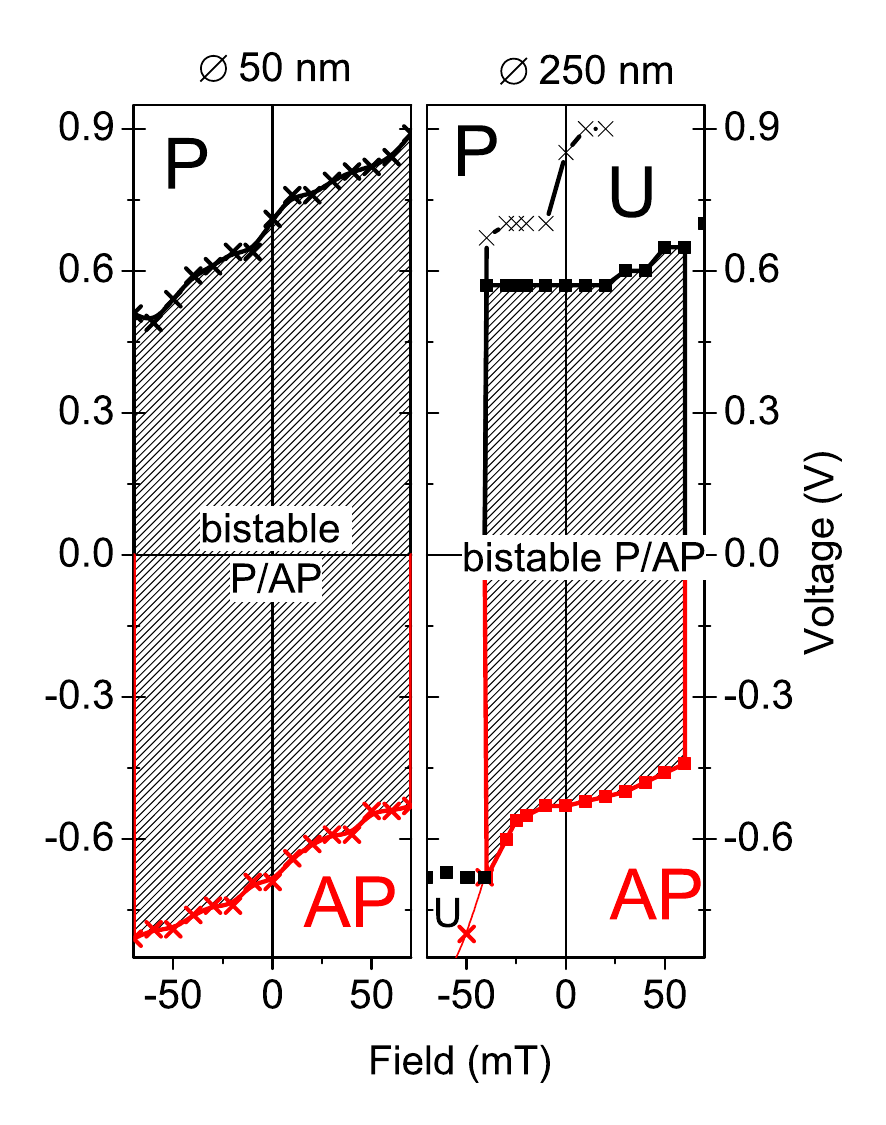}
\caption{(Color online) State diagrams of 50 nm and 250 nm diameter MTJs, as constructued from resistance versus voltage loops recorded at constant fields. U stands for "unknown" state of time-average conductance intermediate between that of P and AP states. } % C4 and C12 devices.
\label{PhaseDiagrams}
\end{figure}

Resistance versus voltage STT loops (Figs.~\ref{RVloops}) have the classical house-like shape \cite{gajek_spin_2012}, with positive voltages favoring the P state and negative voltages favoring the AP state. For junctions of size $2a \leq 110~\mathrm{nm}$, the STT loops involve only the P and AP states at the investigated applied fields. The corresponding phase diagram in the field-voltage region (Fig.~\ref{PhaseDiagrams}) are simple with essentially parallel $\textrm{P} \rightarrow \textrm{AP}$ and $\textrm{AP} \rightarrow \textrm{P}$ branches, in a manner similar to the predictions of the macrospin approximation \cite{bernert_phase_2014}.
For the junctions of sizes $2a \geq 200~\mathrm{nm}$ the STT loops display additional conductance levels that occur in frustrated configurations, i.e. when large fields are applied to hinder a transition favored by STT. These resistance plateaus may correspond to static or dynamic micromagnetic configurations: we emphasize that Fig.~\ref{RVloops}(a) records only the time-average of the conductance of the device in an essentially unknown (U) state that we will determine later. We will see that in these U zones, formerly interpreted as states with static domain wall or steady state oscillations, are far from stationary and involve several intermediate dynamical states that are visited in a telegraph noise manner during the reversal path. Time resolved measurements (section \ref{large}) will reveal that only a fraction of the U states might involve static domain walls. These conductance plateaus are not seen for junctions smaller than typically 150 nm. Just like the coercivity, the STT switching voltage tends to increase for the smallest junctions. The limited life expectancy of the junction when under large voltage stress limits our measurement window to $\pm 1$ V.

%%%%%%%%%%%%%%%%%%%%%%%%%%%%%%%%%%%%%%%%%%%%%%%%%%%%%%%%%%%%
%%%%%%%%%%%%%%%%%%%%%%%%%%%%%%%%%%%%%%%%%%%%%%%%%%%%%%%%%%%%
\section{Spin-torque switching in junctions of diameters in the 150 to 250 nm range} \label{large}
%%%%%%%%%%%%%%%%%%%%%%%%%%%%%%%%%%%%%%%%%%%%%%%%%%%%%%%%%%%%
\subsection{P to AP transition in 250 nm diameter junctions}
\begin{figure*}
\includegraphics[width=16 cm]{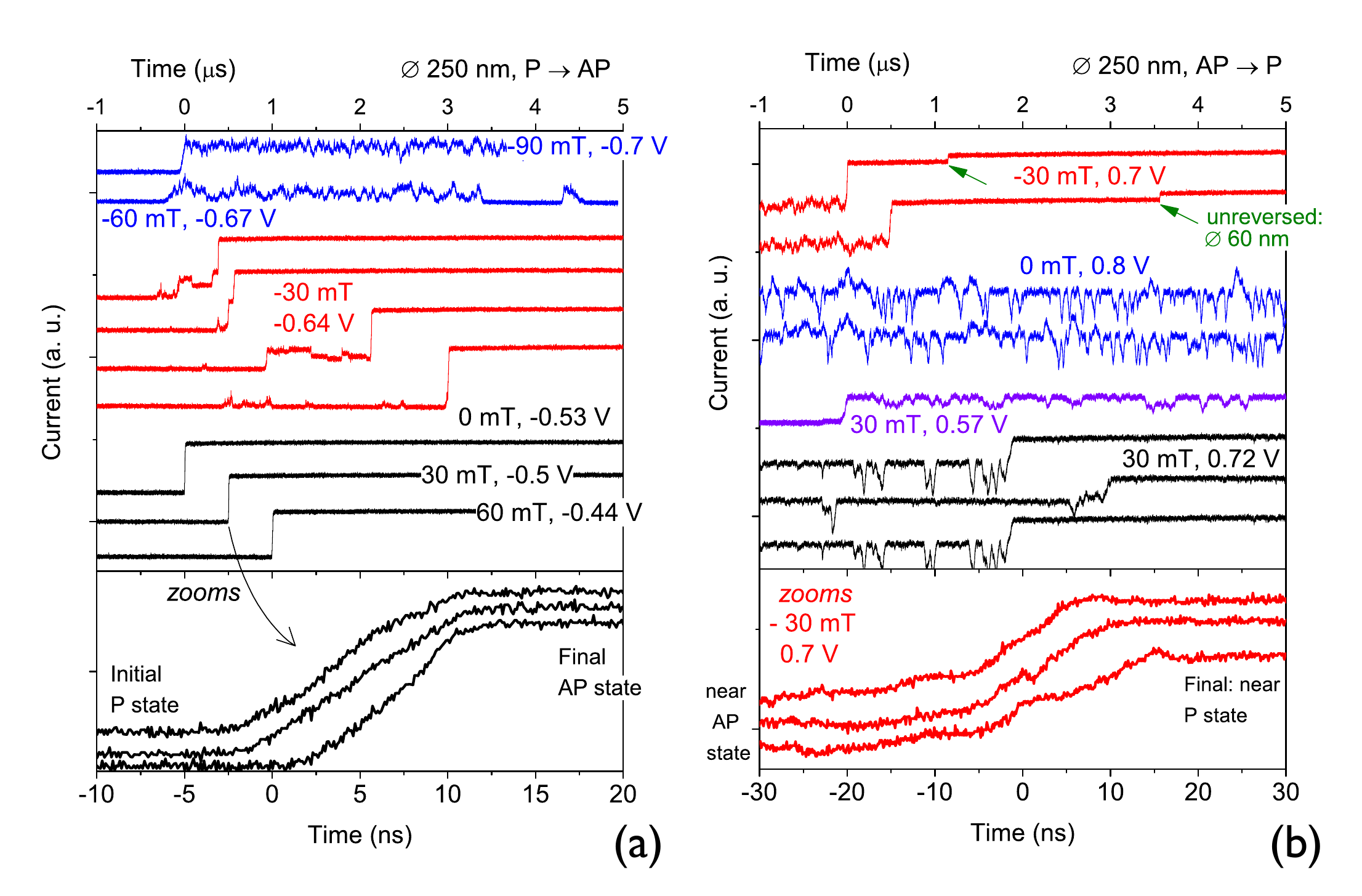}
\caption{(Color online) Time-resolved measurements of sub-thresholds STT-induced $\textrm{P} \rightarrow \textrm{AP}$ (left panels) and $\textrm{AP} \rightarrow \textrm{P}$ (right panels) transitions for junctions of diameter 250 nm. The curves are vertically and horizontally offset for readability. The signal is the voltage delivered to a fixed 50 Ohm load by the (signed) current flowing into the sample, i.e. it has a shape reflecting the opposite (left panels) or similar to (right panels) the conductance evolution. The arrows in the $\textrm{AP} \rightarrow \textrm{P}$ transition indicate the moment at which reversal gets complete. The amplitude of the corresponding signal change is equivalent to the annihilation of a bubble of  diameter 60 nm. } % C12, PAP and APP. 
\label{C12_PAPandAPP}
\end{figure*}

Some representative examples of electrical signatures of the $\textrm{P} \rightarrow \textrm{AP}$ switching in junctions of diameter 250 nm are displayed in Fig.~\ref{C12_PAPandAPP}(a). The switching dynamics seems simple when the applied field and the STT concur to both favor a transition to the AP state. In addition, this requires the minimal voltage. In that case [black traces in Fig.~\ref{C12_PAPandAPP}(a)], the conductance evolve from the microwave quiet P state to microwave quiet AP state in a ramp like manner. The ramps have a rather linear shape with minor event to event variability in the detail of the curvature [bottom panel in Fig.~\ref{C12_PAPandAPP}(a)].  The ramp duration depends on the applied field and it reaches a minimum of 10 ns in near zero applied field.   

The $\textrm{P} \rightarrow \textrm{AP}$ switching is much more complex when the applied field favors the P state and thereby hinders the current-induced reversal. When it weakly opposes the switching, the transition towards AP is sill possible [red curves in Fig.~\ref{C12_PAPandAPP}(a)] at an increased voltage cost. However the irreversible switching event is preceded by repetitive switching attempts, during which the resistance transiently increases to intermediate levels that seem unpredictable. The intermediate levels can be very strongly agitated or can go to rest and yield stable and flat resistance plateaus. When appearing, these stable plateaus in the voltage traces are often followed by irreversible switching happening in a staircase manner, as if a domain wall was moving forward from one parking positions to the next, as observed in other studies on large junctions \cite{devolder_time-resolved_2016}. \\
When the applied field is further increased to impede the $\textrm{P} \rightarrow \textrm{AP}$ transition, the switching does not occur any longer [blue traces in Fig.~\ref{C12_PAPandAPP}(a)] and the onset of dynamics requires an even higher voltage. The system undergoes telegraph noise between the microwave quiet P state and an agitated level. The dwell time of the agitated level increases as the field is strengthened and the frustration is intensified. The time-averaged resistance level is intermediate between the P and AP state and corresponds to the previously observed resistance plateau in the quasi-static R(V) loops (Fig.~\ref{RVloops}) and the resulting U state in the phase diagrams (Fig.~\ref{PhaseDiagrams}).

%%
%	Figure
%%
%

%%%%%%%%%%%%%%%%%%%%%%%%%%%%%%%%%%%%%%%%%%%%%%%%%%%%%%%%%%%%
\subsection{AP to P transition in 250 nm diameter junctions}
Some representative examples of electrical signatures of the $\textrm{AP} \rightarrow \textrm{P}$ switching in junctions of diameter 250 nm are displayed in Fig.~\ref{C12_PAPandAPP}(b). In contrast to the reverse transition, the $\textrm{AP} \rightarrow \textrm{P}$ switching in large junctions is never simple, even when the field and the STT concur to both favor a transition to the P state. In that case [red curves in Fig.~\ref{C12_PAPandAPP}(b)], the destabilization of the initial state (i.e. AP) is evident: prior to the switching event, this destabilization manifests as a very agitated resistance level close to that of AP. Once engaged, the switching is slower than for the reverse transition: it typically takes 25 ns to reach near saturation and it results in creeping ramp-like conductance traces [bottom panel in Fig.~\ref{C12_PAPandAPP}(b)]. Besides, the large resistance step does not correspond to a full switching event; an unreversed area often remains and the switching generally completes only a few microseconds after [see arrows in Fig.~\ref{C12_PAPandAPP}(b)]. 

The situation is even more complex when the applied field favors the AP state and thus hinders the current-induced reversal. When the field is weak or vanishing, a large voltage can force the system to permanently hop between unstable states [blue curves in Fig.~\ref{C12_PAPandAPP}(b)] : the sample can neither stay in the AP state (destabilized by the voltage) nor in the P state (destabilized by the field).  When the applied field favoring the AP state is further increased, the instability of the AP state is apparently attenuated, and AP becomes microwave quiet again. At low voltages close to the onset of the U zone of the diagrams, the system can hop from the microwave quiet AP state to a dynamical state of intermediate average resistance level [violet curve in Fig.~\ref{C12_PAPandAPP}(b)].  Keeping the same field, larger applied voltages (black curves) traversing up the U zone of the state diagram (Fig.~\ref{PhaseDiagrams}) let the system undergo telegraph noise between at least two levels: a state with near AP resistance and an intermediate state. After a variable delay, the system hops irreversibly to a microwave quiet P state and leaves from the U zone the the monostable P zone. When the current is reduced from this U zone, there is a substantial probability of back-hopping to the AP state (not shown).

%%%%%%%%%%%%%%%%%%%%%%%%%%%%%%%%%%%%%%%%%%%%%%%%%%%%%%%%%%%%
\subsection{Summary for junctions of diameters in the 120 to 250 nm range}

To summarize our results on large junctions, the quasi-static resistance versus voltage hysteresis loops indicated that at large fields, the reversal often involves unknown states whose time-averaged resistances are intermediate between P and AP. These states are revealed by time-resolved experiment to be far from stationary and to involve several (if not a continuum of) intermediate dynamical states that are visited in a telegraph noise manner during the reversal path. Only a fraction of these intermediate states are microwave quiet and might involve static domain walls. Applied fields near the compensation point or favoring the AP state can reduce drastically the complexity of the switching path of the P to AP transition, and yield a monotonic and irreversible electrical signatures lasting typically 10 to 15 ns for the largest junctions, and proportionally less small smaller junction diameters. Such a reversal can be interpreted as a  domain wall sweeping irreversibly through the system. Transient domain wall pinning is sometimes seen in the largest junctions. \\
The reverse transition (AP to P) is preceded by a visible strong instability of the AP states. The resistance becomes extremely agitated before switching to P in a path yielding a slow (20 to 50 ns) and irregular increase of the conductance with substantial event-to-event variability for fields near the compensation point. Unreversed bubbles of typical diameter 60 nm can persist a few additional microseconds in the largest junctions.  Fields hindering the AP to P transition yield even more complex dynamics with an intermediate state that is visited in a telegraph noise manner during the reversal path.
At the largest junction size (250 nm), there is no field interval reducing simultaneously the complexity of both the P to AP and the AP to P transition. 

%%%%%%%%%%%%%%%%%%%%%%%%%%%%%%%%%%%%%%%%%%%%%%%%%%%%%%%%%%%%
%%%%%%%%%%%%%%%%%%%%%%%%%%%%%%%%%%%%%%%%%%%%%%%%%%%%%%%%%%%%
\section{Spin-torque switching in medium size junctions (from 80 to 150 nm)} \label{medium}

%%
%	Figure
%%
%
\begin{figure*}
\includegraphics[width=16 cm]{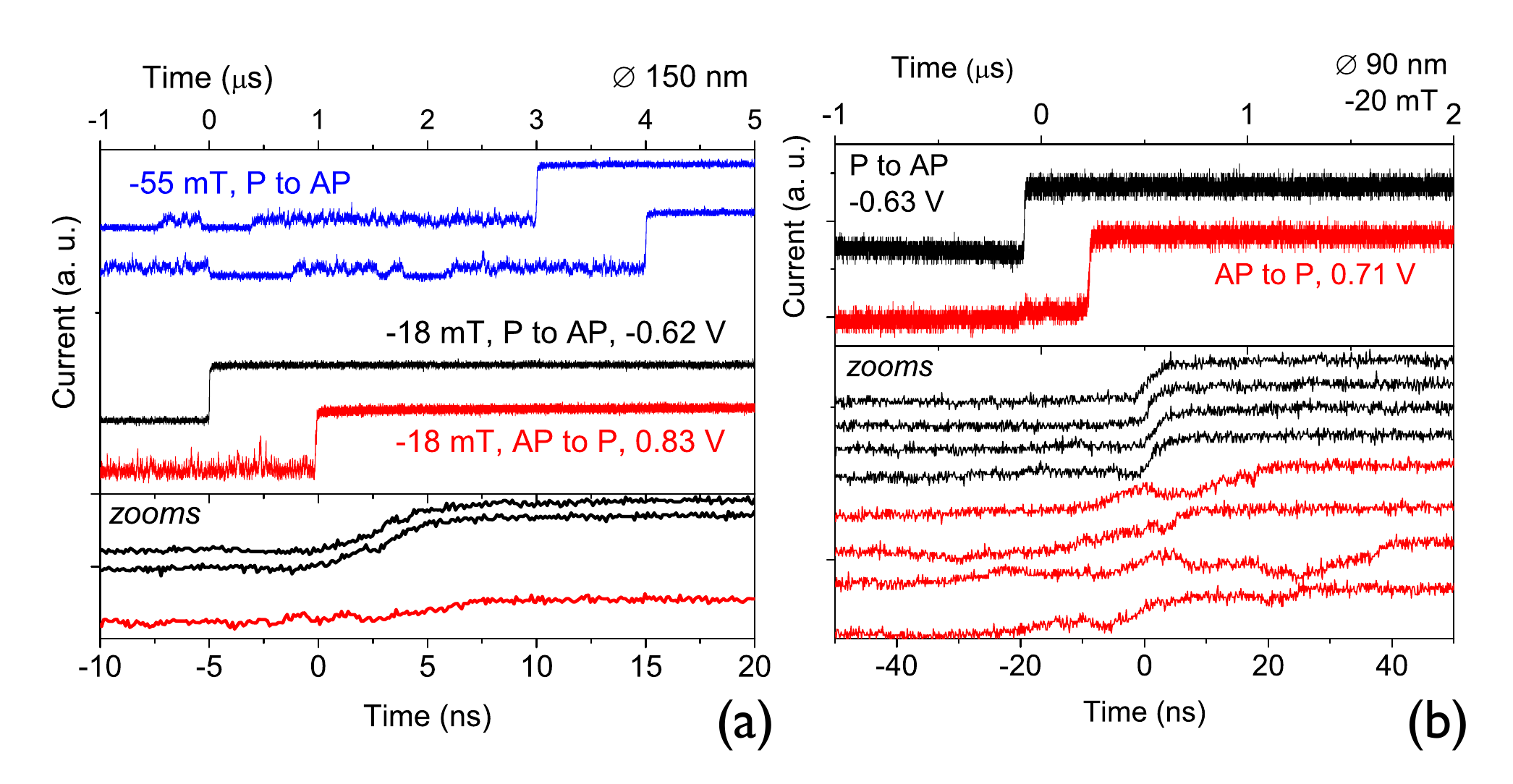}
\caption{(Color online) Time-resolved measurements of sub-thresholds STT-induced $\textrm{AP} \rightarrow \textrm{P}$ transitions for junctions of diameter 150 nm (left panels) and 90 nm (right panels). The curves are offset for readability. The time scales are expended in the bottom panels. } % C10 and C8, PAP and APP.
\label{C10andC8_PAPandAPP}
\end{figure*}

Let us now describe the switching dynamics in medium sized junctions, of diameters typically ranging from 80 nm to 150 nm. As our experimental system measures the current passing through the devices, the signal decreases as the device area, and there is correlatively an apparent increase of noise in the time-resolved traces. This increased instrumental noise [Fig.~\ref{C10andC8_PAPandAPP}(b)] should not be confused with the dynamic states observed formerly in large junctions (Fig.~\ref{C12_PAPandAPP}, blue curves).\\
The situation of junctions of 250 nm diameter was rather extreme since there was no field interval yielding to a simple switching dynamics for both transitions. When reducing the junction sizes, a field interval emerges in which both transitions can occur in an irreversible manner. Fig.~\ref{C10andC8_PAPandAPP} gathers for instance time-resolved transitions in both directions near the field compensation point for two medium size devices.  In line with our prior findings in large junctions, the $\textrm{P} \rightarrow \textrm{AP}$ switching (black curves) is faster and much more reproducible than the $\textrm{AP} \rightarrow \textrm{P}$ (red curves) in medium sizes also. \\%C10 and C8
For neutral fields or fields favoring AP, the $\textrm{P} \rightarrow \textrm{AP}$ switching still exhibits the same simple dynamics leading to ramp-like evolution of the device conductance (black curves). The reversal takes typically 5 ns in 150 nm junctions [Fig.~\ref{C10andC8_PAPandAPP}(a)] 4 ns in 120 nm junctions (not shown), and 3.5 ns in 90 nm junctions [Fig.~\ref{C10andC8_PAPandAPP}(b)], i.e. substantially less than junctions of 250 nm diameters. In that sense, \textit{smaller cells are faster to switch} for the $\textrm{P} \rightarrow \textrm{AP}$ transition. This behavior -- approximate proportional correlation between switching duration and device diameter -- is reminiscent of a domain wall sweeping through the diameter at a velocity that essentially depends on the voltage at which the spin-torque switching occurs \cite{devolder_time-resolved_2016}. 

By contrast, the reverse transitions ($\textrm{AP} \rightarrow \textrm{P}$, red curves in Fig.~\ref{C10andC8_PAPandAPP}) still typically take 20 to 30 ns to complete even in compensated fields conditions. $\textrm{AP} \rightarrow \textrm{P}$ transitions  result in creeping ramp-like conductance traces with a lot of event-to-event variability (Fig.~\ref{C10andC8_PAPandAPP}, red curves). \\ 
Some of the other features observed previously in the large junctions are also still present. In frustrated configurations [for instance field favoring P and STT favoring AP, blue curve in Fig.~\ref{C10andC8_PAPandAPP}(a)], there still exists a dynamical state in which the system hops randomly before the irreversible switching to the AP state. When weakening the frustration by applying only a weak field hindering the current-induced $\textrm{P} \rightarrow \textrm{AP}$ reversal, switching attempts are still seen in a manner very similar to what was displayed in Fig.~\ref{C12_PAPandAPP}(a). Examining the reverse transition ($\textrm{AP} \rightarrow \textrm{P}$) there is still an evident instability of the AP state before $\textrm{AP} \rightarrow \textrm{P}$ switching. This instability leads to permanent resistance agitation for junctions of 150 nm, while it only appears as a resistance increment during a few microseconds before the switching for 90 nm junctions [compare the red curves in the top panels of Fig.~\ref{C10andC8_PAPandAPP}(a) and (b)].

However, some of the features observed previously in the largest junctions seem not to be present in medium size junctions. In particular, the devices not longer stop in microwave quiet states of intermediate resistance levels during the main part of the switching. We tend to think that those states were related to the presence of a domain wall stuck somewhere in the free layer, and that these domain walls are now too wide to be stabilized in junctions of medium sizes. Indications of the presence of unreversed bubbles after an $\textrm{AP} \rightarrow \textrm{P}$ transition were also no longer found in the medium size range.

%%%%%%%%%%%%%%%%%%%%%%%%%%%%%%%%%%%%%%%%%%%%%%%%%%%%%%%%%%%%
%%%%%%%%%%%%%%%%%%%%%%%%%%%%%%%%%%%%%%%%%%%%%%%%%%%%%%%%%%%%
\section{Spin-torque switching in near-50 nm junctions} \label{small}
Let us now describe the switching dynamics in our smallest junctions, that are of diameters 50 and 60 nm. In this size regime the large device resistances have two practical consequences. First, the current passing through the device approaches $100~\mu\textrm{A}$, leading to total signals in the 5 mV range on the $50~\Omega$ oscilloscope, for an apparatus noise of typically 1 mV. This reduced signal-to-noise ratio may lead to difficulties in the identification of dynamical states. Secondly, the small but sizable capacity of 110 fF in parallel with the resistive part of the MTJs leads to a finite bandwidth, such that the measured device response are a convolution of the real device response with the system bandwidth. As demonstrated in Fig.~\ref{C5_PAPandAPPandSETUP}(b), our actual time resolution is slightly below one ns in this size regime. Sub-ns evolutions of the device resistance can no longer be resolved for the smallest junctions. \\

Another consequence of the size reduction is an increase of the switching voltages [compare Fig.~\ref{PhaseDiagrams}(a) and (b)], most probably correlated to the increase of effective anisotropy related to the change of the aspect ratio.

%%%%%%%%%%%%%%%%%%%%%%%%%%%%%%%%%%%%%%%%%%%%%%%%%%%%%%%%%%%%
\subsection{P to AP transition in near-50 nm junctions}
\begin{figure}
\includegraphics[width=9 cm]{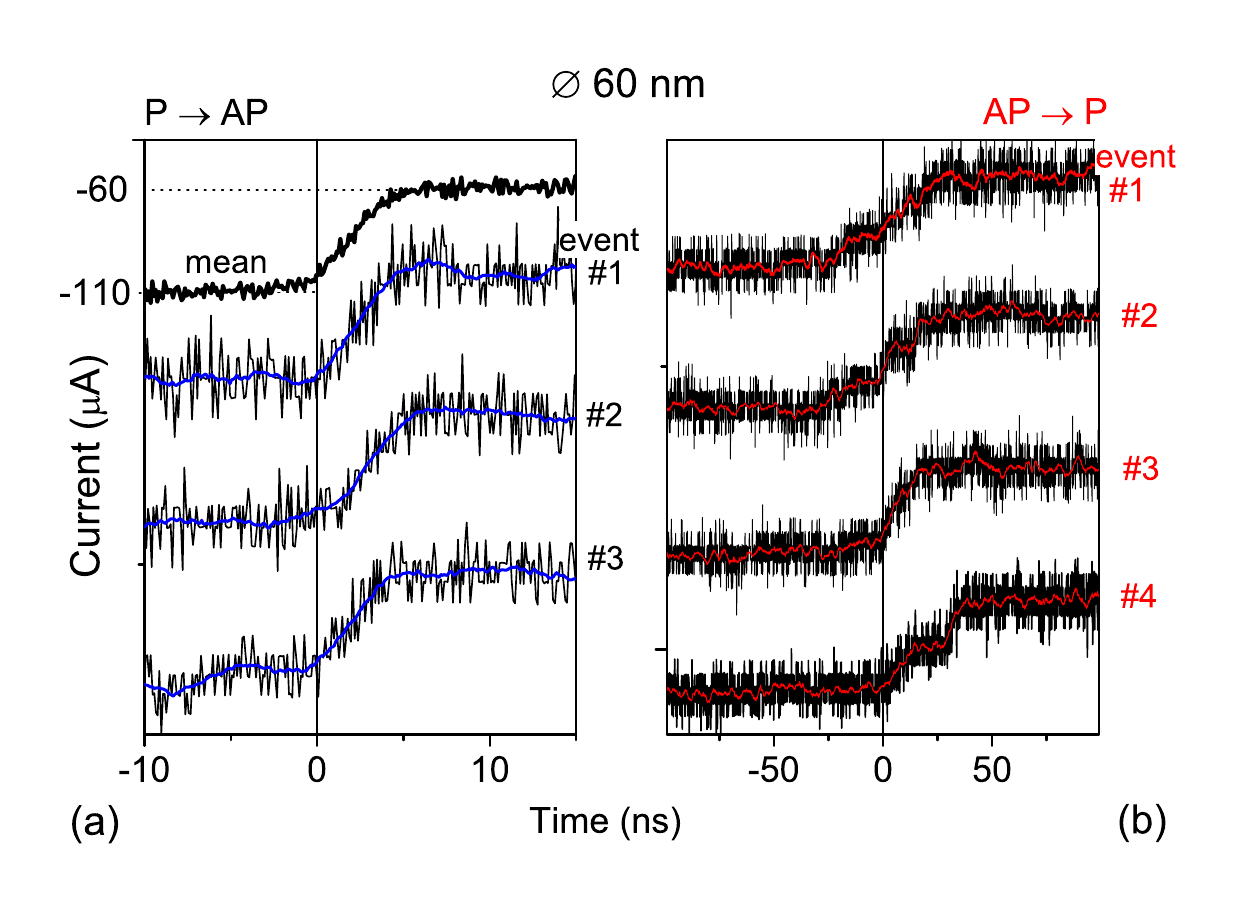}
\caption{(Color online) Time-resolved measurements of sub-thresholds STT-induced transitions for a junction of diameter 60 nm.  The curves are vertically and horizontally offset for readability. The black curves are the raw data, while the colored curves are their counterparts after mathematical low pass filtering of 3 dB cut-off at 1 GHz. (a): examples of $\textrm{P} \rightarrow \textrm{AP}$ transitions. The curve labelled 'mean' is an average of 50 $\textrm{P} \rightarrow \textrm{AP}$ events, meant to evidence the very high reproducibility of the electrical signature of this transition at small junction sizes. (b) examples of $\textrm{AP} \rightarrow \textrm{P}$ transitions. Note the different scales on the time axes. 
 } % C10 and C8, PAP and APP.
\label{C4_PAPandAPP}
\end{figure}

Near 50 nm, the $\textrm{P} \rightarrow \textrm{AP}$ switching proceeds in typically 3 ns, with an irreversible and monotonic increase of the device resistance from the initial P state to the final AP state. In small sized junctions, intermediate state are \textit{never} observed during the $\textrm{P} \rightarrow \textrm{AP}$ transition. The electrical signature of this transition is very reproducible. If there was some event-to-event variability, it would have to be deep sub-ns evolutions, otherwise they would be within reach of our experimental system. \\
To evidence this reproducible character, we have gathered in Fig.~\ref{C4_PAPandAPP} single shot switching events and their mathematically low pass filtered (i.e. smoothed) counterparts next to a mean trace obtained by averaging 50 switching events. The mean trace was constructed by averaging the single event time traces after time-shifting them to align their mid-level switching. The mean trace is very well fitted (not shown) by a function of the type $\mathrm{erf}(t / \tau)$ with $\tau=3.0~\mathrm{ns}$ and resembles the single shot curves, indicating the reproducibility of this transition. We emphasize that this measurement of switching duration is not bandwidth limited.

%%%%%%%%%%%%%%%%%%%%%%%%%%%%%%%%%%%%%%%%%%%%%%%%%%%%%%%%%%%%
\subsection{AP to P transition in near-50 nm junctions}
In contrast, the $\textrm{AP} \rightarrow \textrm{P}$ transition stays complex even for the smallest investigated junctions. On rare occasions, very fast (sub-ns, bandwidth-limited) switching events are seen [see the example in Fig.~\ref{C5_PAPandAPPandSETUP}(b)]. However most of the time, the reversal is much slower and lasts typically 40 to 60 ns. In these cases, after the onset of the reversal, the conductance increases in a staircase-like, monotonic and irreversible manner. During the reversal, the system can stay typically 10 to 20 ns in states with arbitrary intermediate resistance.

%%
%	Figure
%%
%

%%%%%%%%%%%%%%%%%%%%%%%%%%%%%%%%%%%%%%%%%%%%%%%%%%%%%%%%%%%%
%%%%%%%%%%%%%%%%%%%%%%%%%%%%%%%%%%%%%%%%%%%%%%%%%%%%%%%%%%%%
\section{Discussion} \label{discussion}
Let us discuss the possible switching scenario. Until very recently, there was a common belief that PMA systems are faster \cite{worledge_spin_2011} to switch than in-plane magnetized systems, because their gyromagnetic eigenmodes \cite{devolder_performance_2013}, hence the characteristic timescales and attempt frequencies\cite{tomita_unified_2013}  are at much higher frequencies as a result of the strong perpendicular anisotropy. This macrospin-based argument does not seem to apply in the sub-threshold regime that we investigated here. Let us indeed summarize our findings. 

%%%%%%%%%%%%%%%%%%%%%%%%%%%%%%%%%%%%%%%%%%%%%%%%%%%%%%%%%
\subsection{Switching scenario for the $\textrm{P} \rightarrow \textrm{AP}$ transition}

For the $\textrm{P} \rightarrow \textrm{AP}$ transition, there is always a half of the field space in which the reversal can lead to a ramp-like evolution of the device conductance. In the largest cells, one can occasionally observe 'pauses' of several tens of ns during which the device resistance is microwave quiet before it restarts its regular increase. We believe that these pauses are reminiscent of domain wall pinning events during the reversal in large junctions. These pauses (pinning events) were not observed for junction diameters below 150 nm, and generally speaking, downsizing reduces the complexity of the switching to a simple, featureless and more and more reproducible resistance ramp. 

When the $\textrm{P} \rightarrow \textrm{AP}$ reversal happens in this ramp-like manner free of pining events, smaller cells lead to faster switching,  with a switching speed $\tau^{-1}$ that scales approximately like the junction inverse diameter.  Once engaged, the underlying switching process is likely to involve a domain wall that sweeps through the device in a time $\tau = 2a / v$. This would imply a domain wall velocity of typically $v \approx 25 \textrm{~to~} 30 \mathrm{m/s}$ (equivalently: 25 nm/ns) for the voltage at which the reversal occurs (i.e. $\approx -0.6\pm0.1$ V, depending on the field) in our experiments. The field interval in which this reversal mode occurs starts from near the stray field compensation point and extends to the fields favoring the AP state. In practice, it is only limited by our ability to prepare the initial P state by a non-destructive but sufficiently strong voltage. We have not observed features indicative of domain wall propagation in the Walker regime, in contrast to previous observations in elongated (non circular) junctions in which the wall had to travel longer distances \cite{devolder_time-resolved_2016}. 

Despite the reduction of complexity when the junction size is reduced, we have never observed $\textrm{P} \rightarrow \textrm{AP}$ switching events lasting less than 2 ns. This implies that a macrospin-type reversal is \textit{never} happening for the $\textrm{P} \rightarrow \textrm{AP}$ transition, even in junctions as small as  50 nm. We emphasize this point, because the state diagram of Fig.~\ref{PhaseDiagrams}(a) with linear and parallel switching frontiers is sometimes interpreted as an indication of macrospin behavior \cite{timopheev_respective_2015}.

%%%%%%%%%%%%%%%%%%%%%%%%%%%%%%%%%%%%%%%%%%%%%%%%%%%%%%%%%
\subsection{Switching scenario for the $\textrm{AP} \rightarrow \textrm{P}$ transition}
For all investigated junction sizes and applied fields a substantial asymmetry of the switching is found: the $\textrm{AP} \rightarrow \textrm{P}$ transition differs qualitatively from the reverse transition. Sub-ns switching resembling that predicted in the macrospin approximation can occur [Fig.~\ref{C5_PAPandAPPandSETUP}(b)], but this happens with a very low probability, and could only be observed occasionally on the smallest junctions. \\
Most of the time, smaller junctions does not mean faster $\textrm{AP} \rightarrow \textrm{P}$ reversal; the reversal path exhibits a large variability and keeps durations in the 20 to 50 ns duration with no apparent correlation with the junction dimension. The electrical signature of the switching remains complex, with intermediate resistance level and irregular time evolutions even at the smallest dimensions. In that sense, our experimental observations are not compatible with a simple DW sweep from the AP state to the P state. In contrast our results are indicative of strong disorder in the micromagnetic state during the AP to P transition. We emphasize that this strong complexity persist down to 42 nm diameter junctions, despite that this size is only slightly wider than the expected domain wall width. 
%%
%	Figure
%%
%
\begin{figure*}
\includegraphics[width=17 cm]{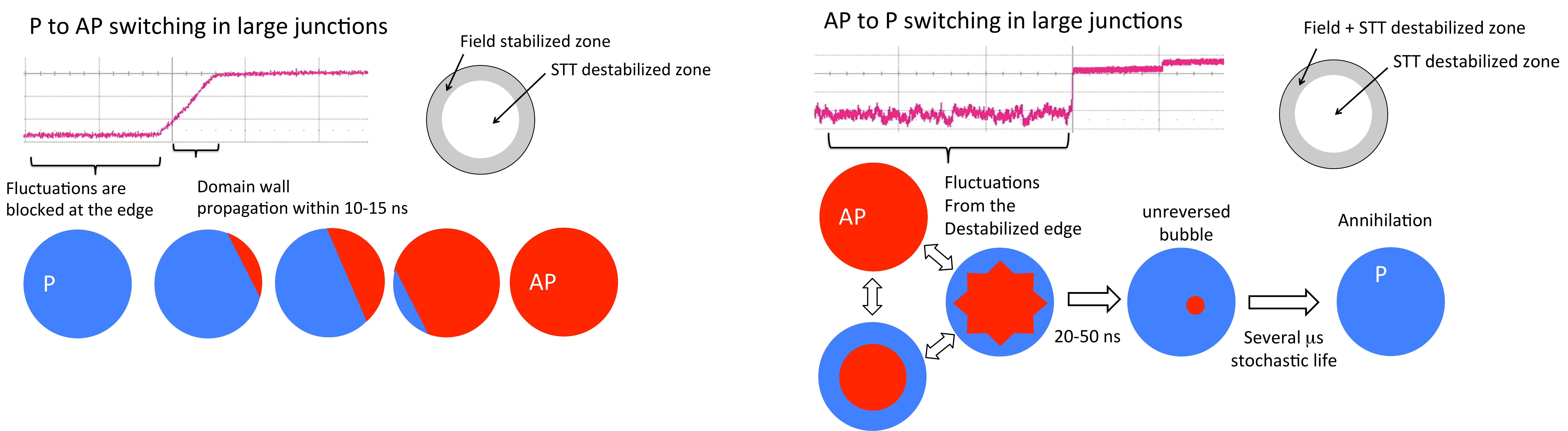}
\caption{(Color online) Proposed spin-torque switching scenario in large (diameter $\geq 100$ nm) junctions in globally compensated field conditions. The $\textrm{P} \rightarrow \textrm{AP}$ transition proceed through easy domain wall motion once a wall has emerged (nucleation) from the edge. This is the limiting step as in the P configuration, the edges of the free layer are stabilized by the dipole field from the closest layer of the reference synthetic antiferomagnet. The $\textrm{AP} \rightarrow \textrm{P}$ proceeds by a complex dynamics involving fluctuations at the junction perimeter, followed by the shrinking of the unreversed domain until the bubble reaches a diameter in the 60 nm range, and finally collapses. The curves are taken from a 250 nm junction.}
\label{scenario}
\end{figure*}

This strong asymmetry between the $\textrm{AP} \rightarrow \textrm{P}$ and the $\textrm{P} \rightarrow \textrm{AP}$ transition is striking and deserves to be commented. A first potential origin is dynamic redistributions of the current density within the nanopillar surface. In metallic nanopillars, this could certainly be considered in the presence of very non uniform time-varying magnetization textures. However in tunnel junctions the spin-torques are proportional to \textit{voltages} \cite{slonczewski_currents_2005}, and the very resistive nature of the MgO oxide in an MTJ ensures that the voltage is spatially uniform across the free layer surface. Current-induced field-like torques, if any, would only shift the behavior along the field axis, which is not observed in our experiments. We thus believe that the spin transport mechanisms can not be responsible for the switching asymmetry. 

On the other end, there is a large influence of the applied field on the overall dynamics. Hence the non uniform \cite{norpoth_straightforward_2008} stray field from the reference system of the tunnel junction may have a drastic influence on the switching and by this induce large asymmetries. This was already conjectured by Gopman et al. from quasi-static experiments \cite{gopman_asymmetric_2012, gopman_bimodal_2014}. Despite the compensation of the reference layer stray fields which is achieved though engineering of the synthetic antiferromagnetic reference layer system \cite{gottwald_scalable_2015}, there remains inevitably some non uniform stray field. When globally compensated, the stray field of a synthetic antiferromagnetic reference system has opposite polarities near the edge and at the center of the device \cite{devolder_time-resolved_2016}. 

This implies that the reversal is impeded near the edges during a $\textrm{P} \rightarrow \textrm{AP}$ transition (i.e. in the P configuration, the edges of the free layer are stabilized by the dipole field). Once a domain wall, (which can only enter from the edge from topological reasons) has crossed the near edge area for the so-called in nucleation step, there is no reason for the wall not to sweep easily through the whole device. This leads to irreversible domain wall sweep in this $\textrm{P} \rightarrow \textrm{AP}$ transition, as sketched in Fig.~\ref{scenario}. \\
In contrast, when in the AP state, the reference layer stray field favors the reversal near the edges and impede the reversal near the center. In $\textrm{AP} \rightarrow \textrm{P}$ transitions, the domain walls (or other kinds of non uniform fluctuations) can easily enter from the edges but cannot penetrate further to the inner part of the free layer disk. This can be the reason why the initial state is very noisy before the $\textrm{AP} \rightarrow \textrm{P}$ transitions for the junctions whose sizes make it conceivable to host strong magnetization non uniformities. The fluctuations can finally penetrate to the inner part of the free layer, and converge either to saturation or to form a tiny unreversed bubble of typical diameter 60 nm. When created, the bubble annihilation can last a stochastic time, varying in practice in the range of several  microseconds, as sketched in Fig.~\ref{scenario}.

The generalization of the two switching scenario sketched in Fig.~\ref{scenario} to the simpler case of small junctions is straightforward. The duration of the $\textrm{P} \rightarrow \textrm{AP}$ transition is shrunk as the junction diameter, simply because the wall velocity is determined by the current density~\cite{devolder_time-resolved_2016} which is almost independent from the junction size. In contrast, the $\textrm{AP} \rightarrow \textrm{P}$ transition is affected qualitatively by the change of dimension for two reasons. First, because shrinking the junction makes it unable to host a stable bubble whose size can not be compressed below a few Bloch parameters. Second, because the size of the destabilized annular zone (gray color in Fig.~\ref{scenario}) gets too small to host substantial fluctuations; an apparently stable AP state is thus recovered prior to the switching event, but the switching is still very complex, as indeed observed in our smallest junctions.

%%%%%%%%%%%%%%%%%%%
%%%%%%%%%%%%%%%%%%%	
\section{Conclusion}
In summary, we have studied spin-transfer-torque (STT)-induced switching in perpendicularly magnetized tunnel junctions (pMTJ) of sizes from 50 to 250 nm. The studied reversal regime is the sub-threshold thermally activated reversal.\\ At the largest junction size (diameter 250 nm), the quasi-static resistance versus voltage hysteresis loops indicate that the reversal often involves states whose time-averaged resistance is intermediate between P and AP. These states, formerly interpreted as states with static domain wall, are revealed by time-resolved experiment to be far from stationary and to involve several intermediate dynamical states that are visited in a telegraph noise manner during the reversal path. Only a fraction of the intermediate state might involve static domain walls. Specific applied fields can reduce drastically the complexity of the switching path and yield electrical signature indicative of a monotonic and irreversible reversal lasting typically 10 to 15 ns, that can be interpreted as a  domain wall sweeping through the system at 16 to 25 nm/ns.

The complexity of the switching is gradually reduced when the junctions are downsized. From 250 to 150 nm device diameter, when away from optimal applied field conditions, the P to AP switching is preceded by repetitive switching attempts, during which the resistance is transiently increased from the microwave quiet P state. At 50 nm, the P to AP switching proceeds in typically 2-3 ns, with an irreversible and monotonic increase of the device resistance from quiet P state to quiet AP state, that we interpret as a simple domain wall propagation process, again sweeping through the system at typically 25 nm/ns.

Concerning the AP to P transition, in devices of 250 to 150 nm diameter, the AP to P switching is preceded by a strong instability of the AP states, whose resistance becomes extremely agitated before switching to AP. The AP to P reversal is slow (20 to 50 ns) and yields irregular resistance traces with very substantial event-to-event variability. Some unreversed bubbles of typical diameter 60 nm can persist a few additional microseconds in the largest junctions. Below 60 nm, no instability of the initial AP stat is detected priori to switching but the switching stays slow (20 to 50 ns) and still yields the same irregular resistance traces with tremendous event-to-event variability. In the smallest junctions (50 nm) we occasionally observe much faster (sub-1 ns) AP to P switching events, possibly consistent with a macrospin process. 

Our results indicate that the reversal path is richer than commonly thought. In contrast to in-plane magnetized systems,  the variability of thermally activated reversal is not restricted to stochastic variations of a incubation delays before the onset of reversal: several reversal paths are often possible even at the smallest dimensions. Besides, the non uniform nature of the magnetic response seems still present at the nanoscale in the AP to P transition; this has implications for the understanding of the numerous systems where spatial coherence of the spin system is crucial. In particular, our results call for a revisit of the modeling of switching dynamics in STT-operated systems of perpendicular anisotropy. \\
%%%%%%%%%%%%%%%%%%%
%%%%%%%%%%%%%%%%%%%	
This work was supported by the Samsung Global MRAM Innovation Program.

%\bibliography{bib.bib}
%merlin.mbs apsrev4-1.bst 2010-07-25 4.21a (PWD, AO, DPC) hacked
%Control: key (0)
%Control: author (8) initials jnrlst
%Control: editor formatted (1) identically to author
%Control: production of article title (-1) disabled
%Control: page (0) single
%Control: year (1) truncated
%Control: production of eprint (0) enabled
%

\end{document}